\documentclass[12pt,aps,nofootinbib,superscriptaddress,showpacs]{revtex4}
\usepackage{epsfig}
\usepackage{bm}
\usepackage{color}
\usepackage{amsmath}
\begin{document}
\title{R-Matrix theory with level-dependent boundary condition parameters}
\author{Tae-Sun Park}
\affiliation{
Center for Exotic Nuclear Studies, Institute for Basic Science, Daejeon 34126, Korea}
\date{\today}

\def\be{\begin{eqnarray}}
\def\ee{\end{eqnarray}}
\def\vI{{\bm V}}
\def\vL{{\bm L}}
\def\calS{{{\cal S}}}
\def\calR{{\cal R}}
\def\calRB{\calR^B}
\def\calB{\cal B}
\def\calN{\cal N}
\def\calE{\cal E}
\def\bcalE{\bar {\cal E}}
\def\vphi{\varphi}
\def\J{J}

\begin{abstract}
I present a new formalism of the R-matrix theory where 
the formal parameters for the
resonance energies and widths
are identical to the observed values.
By allowing
the boundary condition parameters
to vary from level to level,
the freedom required 
to adjust the formal parameters for the pole positions
to the observed values is obtained.
The basis 
of the resulting theory
becomes non-orthogonal,
and I describe the procedure to construct
a consistent R-matrix theory with
such a non-orthogonal basis. 
And by adjusting the normalization of the states that form the basis,
the formal parameters for the reduced decay widths
also become the same as those observed,
leaving no formal parameters that are different from the observed ones.
A demonstration of the developed theory 
to the elastic 
${}^{12}\text{C}+p$ scattering data
is presented.

\end{abstract}
\maketitle

\section{Introduction}

The R-matrix theory~\cite{kapur,we}
introduced 
by Wigner and Eisenbud~\cite{we}
and by Kapur and Peierls~\cite{kapur}
is 
an extremely powerful and 
indispensable tool
for describing nuclear reactions~\cite{lane,descouvemont},
in which
collision matrices 
are
described in terms of the positions
and widths of the resonances.
The values of the resonance parameters used in the R-matrix theory
(which are referred to as the {\em formal} parameters)
are, however, 
different
from the {\em observed} values,
and the conversion between the two sets of parameters 
is non-linear.
%
%

Having two different sets of parameters
complicates
both theoretical and computational studies of 
the theory quite substantially,
see Ref.~\cite{descouvemont} for a detailed discussion.
A potential improvement is to derive
an R-matrix theory where the formal parameters coincide with the observed values
by a reformulation of the theory,
such that 
unnecessary
confusion and complications can be removed.

In this paper, I address this issue by presenting a new R-matrix theory  
where there is no distinction between the two sets of parameters.
The key steps of the developed formulation can be summarized as follows.
When constructing the conventional R-matrix theory,
the basis 
for the Hilbert space
defined 
in the {\em internal} region 
of the theory
is required
to be orthogonal.
Since the domain of the basis is 
not the whole configuration space
but limited only
to the internal region,
the seemingly innocent orthogonality condition
imposes quite a strict constraint on the 
states that form the basis:
The boundary condition parameters
that define the basis states should be level-independent
in order to satisfy the orthogonality condition
(see the text for more detailed explanation).
This constraint is
in fact
responsible for 
the discrepancies between
the formal parameters 
and
the observed ones.
%
%
However, it is possible to construct
a consistent R-matrix theory
with a non-orthogonal basis.
It turns out that
releasing the orthogonality condition
provides the freedom 
to have 
level-dependent
boundary conditions,
with which 
the {\em formal} parameters 
can be equivalent 
to the {\em observed} values.
To be more specific,
if 
one adjusts 
the boundary condition parameters $B_{\lambda c}$ 
to 
the shift factor
$S_c$ at $E_\lambda$
(the pole-position of level $\lambda$), 
the formal parameters for the pole positions coincide with the observed ones.
Furthermore,
there is additional freedom associated with
the normalization of the basis states.
By selecting the normalization given in Eq.(\ref{Kll}),
the formal parameters for reduced widths also become 
the same as observed.
As a result, there are no formal parameters that are different from the observed values
in the present theory,
achieving the desired goal.

The new formalism shares many features in common with 
previous studies.
For example,
the very original work by Kapur and Peierls~\cite{kapur}, 
has an energy-dependent
boundary condition.
Schemes with level-dependent boundary condition parameters 
were introduced 
by Barker~\cite{barker}
and by Azuma et al.~\cite{azuma}.
Angulo and Descouvemont developed a formalism~\cite{angulo}
where there are no level shifts, 
which is however applicable only to single-channel cases.
In particular, this work is 
rather similar to the 
work of Brune~\cite{brune},
where 
an alternative parametrization was introduced
to use the observed 
pole locations 
as inputs of the R-matrix theory, developing
a transformation scheme between the formal and the observed parameters.
In this work, 
the theory is formulated in such a way that
all the formal parameters 
are directly equivalent to
the observed values.

In Section II, I briefly review the R-matrix theory,
explaining how the orthogonality condition can be released
in a consistent manner to allow the boundary condition parameters to
be level-dependent.
In Section III,
I then describe the procedure
to align the formal parameters to the observed values.
In Section VI, a demonstration of the developed theory is made for the 
${}^{12}\text{C}+p$ elastic scattering reaction.
Section V is devoted to discussions.

\section{Formalism for the extended boundary condition}

I begin with a brief review of the R-matrix theory
described in Lane and Thomas (LT)~\cite{lane}.
%
%
%
In the R-matrix theory, 
a nuclear system is described in terms of
channels that consist of two subsystems,
$\alpha_1=\left\{Z_1,\,A_1,\,I_1\right\}$ and
$\alpha_2=\left\{Z_2,\,A_2,\,I_2\right\}$,
where
$Z_i$, $A_i$ and $I_i$ are the
proton number, mass number and spin of the $i$-th subsystem,
respectively.
The quantum state of 
a channel $c$
might be denoted as
$c=\left\{\alpha, (I_1,I_2) s, \ell; J J_z\right\}$,
where 
$\alpha=\alpha_1\otimes\alpha_2$ is the partition index,
$s$ is the channel spin ($\vec{s} = \vec{I}_1 + \vec{I}_2$),
$\ell$ is the relative orbital angular momentum,
$J$ is the total angular momentum ($\vec{J}=\vec{s}+\vec{\ell}$)
and $J_z$ is its projection~\cite{descouvemont}.
%
The Hamiltonian of the system in the center-of-mass frame then reads
\be
H= - \frac{\hbar^2}{2 M_c} \nabla_{\vec{r}_c}^2 + V_c (\vec{r}_c) + H_{\alpha_1} + H_{\alpha_2},
\label{H}\ee
where $\vec{r}_c$ and $M_c$ are the relative position vector and the reduced mass of channel $c$, respectively,
and
$H_{\alpha_i}$ is the Hamiltonian for the internal energy of the $i$-th subsystem.
The wavefunction 
of the system
that satisfies the Schr\"odinger equation at energy $E$,
$H\Psi = E \Psi$
can be written as
\be
\Psi = \sum_c \vphi_c \, u_{c}(r_c),
\ee
where 
$u_c(r_c)$ is 
the radial function of channel $c$.
The spinors and the angular dependence of the subsystems are embodied
in 
$\vphi_c = \frac{1}{r_c} \left[\left[\vphi_{\alpha_1} \otimes
\vphi_{\alpha_2}\right] \otimes i^\ell Y_{\ell}(\hat{r}_c)\right]_{J J_z}$~\cite{descouvemont}.
The radial space of each channel
is divided into two parts: the internal region ($r_c \leq a_c$)
and the external region ($r_c > a_c$),
where $a_c$ is the {\em channel radius} that defines the surface 
between the two regions.
$\vphi_c$
is assumed to be orthonormal when integrated on the channel surface,
\be
\int \vphi_c^* \vphi_{c'} \, d\calS = \delta_{c c'},
\label{phiphi}\ee
where $\calS$ is the surface defined by $r_c=a_c$.
Here and hereafter, unless stated otherwise,
I follow the notation given in LT.

The wave function in the internal region
is expanded 
as a linear combination 
of the basis states 
$X_\lambda$,
\be 
\Psi = \sum_\lambda C_\lambda X_\lambda,
\label{PsiCX}\ee
where
$C_\lambda$ are $E$-dependent coefficients.
The basis states
\be
X_\lambda = \sum_c \vphi_c \, u_{\lambda c}(r_c)
\label{Xlambda}\ee
are defined by the eigenvalue equation
\be
H X_\lambda = E_\lambda X_\lambda\,.
\label{eigen}\ee
with a boundary condition imposed at the channel surface,
which will be discussed soon.

Since the basis states
are relevant only in the internal region,
I define their inner products as\footnote{\protect
By making use of Eqs.(\ref{Xlambda}) and (\ref{phiphi}),
$\J_{\lambda \lambda'}$ can also be represented as
$$\J_{\lambda \lambda'} 
= \sum_c \int_0^{a_c}\!\! d r_c \, u_{\lambda c}^*(r_c)
u_{\lambda' c}(r_c).$$}
\be
\J_{\lambda \lambda'} 
&\equiv& \int_\tau X_{\lambda}^* X_{\lambda'}\,d\tau,
\ee
where
$\int_\tau \, d\tau$ denotes
the volume integral limited only to
the internal region.
%
The orthogonality condition of the basis
then corresponds to 
have
$J_{\lambda\lambda'}=0$ for $\lambda\neq \lambda'$.
%
%
$J_{\lambda\lambda'}$ can be evaluated by 
the following steps:
If one multiplies
Eq.(\ref{eigen}) by $X_{\lambda'}^*$ 
from the left
and integrate 
it in the internal region, and then subtract it with
interchanging $\lambda'$ and $\lambda$,
one obtains
%
\be
(E_{\lambda}-E_{\lambda'}) \int_\tau X_{\lambda}^* X_{\lambda'} \, d\tau
&=& \int_\tau \left[
   \left(H X_{\lambda}\right)^* X_{\lambda'} 
   - X_{\lambda}^* \left(H X_{\lambda'}\right)\right]\,d\tau
\nonumber \\
&=& - \sum_c \frac{\hbar^2}{2 M_c} \left(
      u_{\lambda' c} \frac{d u_{\lambda c}}{dr}
      - u_{\lambda c} \frac{d u_{\lambda' c}}{dr}
      \right)_{r=a_c},
\label{EsubE}\ee
where I have inserted
the Hamilton given in Eq.(\ref{H})
at the last step.
Here and hereafter, 
I limit myself
to the cases where 
the nuclear potential is hermitian
and the radial functions are real,
$u_{\lambda c}(r)^* = u_{\lambda c}(r)$.
Diving the above equation by $E_{\lambda}-E_{\lambda'}$, 
one is 
then led to\footnote{\protect
Here I assume that for a given spin and parity,
$X_\lambda$ is non-degenerate, and thus 
$\lambda\neq \lambda'$ implies $E_\lambda \neq E_{\lambda'}$.
The degenerate levels with the same level-energy, if any, 
can be merged into a single level, as discussed in Ref.~\cite{brune}.}
\be
\J_{\lambda \lambda'} 
&=& 
- \frac{1}{E_{\lambda}-E_{\lambda'}} 
 \sum_c \gamma_{\lambda c} 
\left( B_{\lambda c}  -  B_{\lambda' c} \right) 
 \gamma_{\lambda' c}, 
 \ \ \ \text{for}\ \lambda \neq \lambda',
\label{Kgen}
\ee
where 
\be
\gamma_{\lambda c} &\equiv& \sqrt{\frac{\hbar^2}{2 M_c a_c}}\, u_{\lambda}(a_c),
\\
B_{\lambda c} &\equiv& \left.\frac{a_c}{u_{\lambda c}(a_c)} 
\frac{d u_{\lambda c}(r)}{dr}\right|_{r=a_c}.
\ee
%
From Eq.(\ref{Kgen}), it is clear that
the orthogonality condition for a general
multi-level and multi-channel case
can be guaranteed only when
the boundary condition is level-independent,
$B_{\lambda c} = B_c$,
as is demanded in the 
conventional R-matrix theory.
%
%
However, the orthogonality is not a necessary condition
for the basis of a consistent R-matrix theory.
If one does not adhere to it,
as I explain below,
one is granted additional freedom to have level-dependent
boundary condition parameters
that can be used to remove the gap between the formal
parameter set
and the observed parameter set.
%

I now describe how a new R-matrix theory
can be built with a non-orthogonal basis.
From Eqs.(\ref{Kgen}) and (\ref{PsiCX}), 
the coefficients $C_\lambda$ read
\be
C_{\lambda'} = \sum_{\lambda} \left(\J^{-1}\right)_{\lambda' \lambda}
\int_\tau X_{\lambda} \Psi\,d\tau ,
\label{Clambda}\ee
and the integral in the above equation
can be evaluated by
inserting $\Psi$ in place of $X_{\lambda'}$ in Eq.(\ref{EsubE}),
\be
\int_\tau X_{\lambda} \Psi\,d\tau
= \frac{1}{E_{\lambda}-E} 
 \sum_c \gamma_{\lambda c}
\sqrt{\frac{\hbar^2}{2 M_c a_c}} 
\left(a_c \frac{d u_c(r)}{dr} - B_{\lambda c} u_c(a_c)\right)_{r=a_c}.
\ee
Insertion of the resulting coefficients into
Eq.(\ref{PsiCX}) gives the following equation,
\be
u_{c'}(a_{c'}) &=& \sum_{\lambda'} C_{\lambda'} u_{\lambda' c'}(a_{c'})
\nonumber \\
&=& \sum_c \sqrt{\frac{M_{c'} a_{c'}}{M_{c} a_{c}}}
\left(\calR_{c'c} \, a_c \frac{d u_c(r)}{dr}
- \calRB_{c' c} u_c(a_c)\right)_{r=a_c}
\label{uin}\ee
with
\be
\calR_{c'c} &\equiv&
\sum_{\lambda',\lambda} \gamma_{\lambda' c'}
\left(\J^{-1}\right)_{\lambda' \lambda} \frac{1}{E_\lambda-E} \gamma_{\lambda c},
\label{calR}
\\
\calRB_{c'c} &\equiv&
\sum_{\lambda',\lambda} \gamma_{\lambda' c'}
\left(\J^{-1}\right)_{\lambda' \lambda} 
\frac{1}{E_\lambda-E} \gamma_{\lambda c}\, B_{\lambda c}.
\label{calRB}
\ee
%
On the other hand,
the radial wave functions in the external region
can be written
analytically,
because the channel radius $a_c$ is assumed to be large enough so that
all the nuclear forces between the two subsystems vanish
and only the Coulomb interaction remains,
\be
u_{c'}(r) = \frac{1}{\sqrt{v_{c'}}} \sum_c
\left[ I_{c'}(r) \delta_{c'c} - O_{c'}(r) U_{c'c} \right] y_c,
\ \ \ r_c \geq a_c,
\label{uout}\ee
where $y_c$ are coefficients,
$I_c(r)$ and $O_c(r)$ are the incoming and outgoing radial wave functions, respectively,
$U$ is the collision matrix, and 
$v_c = \sqrt{2 |E_c|/M_c}$ are the relative velocities.
The collision matrix can be obtained
by requiring that
the logarithmic derivatives of the radial functions
on the channel surface
resulting from Eq.(\ref{uout}) 
should be equal to 
the derivatives derived from Eq.(\ref{uin}), 
\be
U_{c'c}= \Omega_{c'} \left(\delta_{c'c} + 2i \sqrt{P_{c'}} 
      \left(\left[1-\calR (S_c+iP_c) + \calRB\right]^{-1} R\right)_{c'c} \sqrt{P_c}
      \right) \Omega_c,
\label{URR}\ee
where $\Omega_c= \sqrt{I_c/O_c}$,
and the shift ($S_c$) and penetration ($P_c$) factors are
the real and imaginary parts of 
the logarithmic derivative of the outgoing wavefunction
on the channel surface, respectively,
\be
\left. r (\partial O_c/\partial r)/O_c\right|_{r=a_c}
= S_c + i P_c.
\ee


In this context,
it is convenient to represent
the collision matrix in terms of the
so-called $A$-matrix,
which is defined by
\be
\left(\left[1-\calR (S+iP) + \calRB\right]^{-1} R \right)_{c'c}
= \sum_{\lambda',\lambda} \gamma_{\lambda' c'} A_{\lambda'\lambda} \gamma_{\lambda c},
\ee
or, equivalently,
\be
\left(A^{-1}\right)_{\lambda \lambda'} &=&
{\bcalE}(E)_{\lambda \lambda'} 
 - i \sum_c \gamma_{\lambda c} \gamma_{\lambda' c} P_c(E),
\label{AinvE}
\ee
where ${\bcalE}(E)$ is the real part of $A(E)^{-1}$,
\be
{\bcalE}(E)_{\lambda \lambda'} &=&
(E_\lambda -E) \J_{\lambda \lambda'} 
- \sum_c \gamma_{\lambda c} \left[ S_c(E) - B_{\lambda c}
\right] \gamma_{\lambda' c}.
\label{calE}
\ee
The collision matrix with this $A$-matrix reads
\be
U_{c'c}= \Omega_{c'} \left(\delta_{c'c} + 2i \sqrt{P_{c'}} 
 \sum_{\lambda',\lambda} \gamma_{\lambda' c'} A_{\lambda'\lambda} \gamma_{\lambda c}
      \sqrt{P_c} \right) \Omega_c .
\label{UA}\ee
Using the above Eqs.(\ref{AinvE},\ref{calE},\ref{UA}),
one can thus construct the collision matrix
with the general inner products of the basis states
given in Eq.(\ref{Kgen}),
and the basis no longer needs to be orthogonal.
The values of the level-dependent boundary condition
parameters $B_{\lambda c}$ 
and the diagonal elements $J_{\lambda\lambda}$ 
should then be
determined,
which will be discussed in the next section.

\section{Determination of the boundary condition parameters}


So far, I have shown that
releasing the orthogonality condition of the basis states
allows the boundary condition parameters $B_{\lambda c}$ to depend on the level.
%
This section describes 
how to utilize
this additional freedom associated with the level-dependence
to make the {\em formal} parameters coincide with the observed ones.

Consider first the {\em observed} pole-positions of resonances,
$E_\lambda^{\text{obs}}$.
%
%
The precise definition of the pole-position may be ambiguous,
and I adopt the convention of Ref.~\cite{brune}
where $E_\lambda^{\text{obs}}$ are defined 
as the zeroes of the determinant of the real part of the inverse of the 
$A$-matrix,
or, equivalently,
the solutions of the secular equation
\be
\det\bcalE(E) = 0.
\label{secular}\ee
%
The aim 
of equalizing the observed pole positions with the formal parameters 
\be
E_\lambda^{\text{obs}} = E_\lambda
\label{EE}\ee
can be achieved
if one sets the boundary condition parameters
$B_{\lambda c}$ to be the shift factor at $E=E_\lambda$,
\be
B_{\lambda c}= S_c(E_\lambda).
\label{Blc}\ee
%
This can be seen by simply noting that
${\bcalE}(E_\lambda)_{\lambda \lambda'}$
vanishes if one inserts Eq.(\ref{Blc}) into Eq.(\ref{calE}).
That is, 
for any $\lambda$,
the entire $\lambda$-th row of the matrix
${\bcalE}(E_\lambda)$ vanishes,
which in turn makes
$E_\lambda$ the solution of Eq.(\ref{secular}).
This proves that 
the formal parameter $E_\lambda$ is equal to the {\em observed} $E_\lambda^{\text{obs}}$.


From Eqs.(\ref{AinvE},\ref{calE},\ref{UA}), it is not difficult to see that
the collision matrix $U$ is invariant under the following transformation,
\be
\J_{\lambda\lambda} &\to& 
\J_{\lambda\lambda}^{\text{new}},
\nonumber \\
\gamma_{\lambda c} &\to& 
\gamma_{\lambda c}^{\text{new}}  =
\sqrt{\frac{\J_{\lambda\lambda}^{\text{new}}}{\J_{\lambda\lambda}}} \,\gamma_{\lambda c}.
\ee
The normalization of $\gamma_{\lambda c}$
is thus determined by 
the values of $\J_{\lambda\lambda}$,
which are not yet determined.
The off-diagonal elements of $J$
are given in  Eq.(\ref{Kgen}).
A natural extrapolation 
to the diagonal cases 
would be
to take the limit $E_{\lambda'}\to E_\lambda$
of the equation,
which results in, with Eq.(\ref{Blc}),
\be
\J_{\lambda\lambda}= 1-\sum_c \gamma_{\lambda c}^2\, \left.
\frac{d S_c(E)}{d E}\right|_{E=E_\lambda}.
\label{Kll}\ee
Insertion of the above equation into
Eq.(\ref{calE}) yields
\be
{\bcalE}(E)_{\lambda \lambda'} =
\begin{cases}
E_\lambda - E - \sum_c \gamma_{\lambda c}^2
\left[ S_c(E) - S_{\lambda c}
+ (E_\lambda-E) S_{\lambda c}'
\right],
   & \text{for } \lambda'=\lambda,\\
- \sum_c \gamma_{\lambda c} \gamma_{\lambda' c}
\left[ S_c(E)
+  \frac{(E_{\lambda'}-E) S_{\lambda c} -(E_{\lambda}-E) S_{\lambda' c})}{
      E_\lambda- E_{\lambda'}}
\right],
   & \text{for } \lambda'\neq \lambda,\\
\end{cases}
\label{Ainv2}\ee
where $S_{\lambda c} \equiv S_c(E_\lambda)$.
%
%

I now consider the consequence of Eq.(\ref{Kll})
on the {\em observed} widths of the resonances,
which are usually defined with the ``one-level approximation".
With this approximation,
the collision matrix 
reads (LT)
%
\be
U_{c'c}\simeq \Omega_{c'} \left(\delta_{c'c} + i 
 \frac{ \sqrt{\Gamma_{\lambda c'}} \sqrt{\Gamma_{\lambda c}}}{E_\lambda-E- \frac{i}{2} \sum_{c''}\Gamma_{\lambda c''}}
      \right) \Omega_c. 
\label{UA0}\ee
And the 
{\em observed} reduced widths $\gamma_{\lambda c}^{\text{obs}}$
are defined by~\cite{descouvemont}
\be
\Gamma_{\lambda c}(E_\lambda) = 2 P_c(E_\lambda) 
   \left.\gamma_{\lambda c}^{\text{obs}}\right.^2.
\ee
On the other hand, 
if one inserts Eq.(\ref{Ainv2}) into Eq.(\ref{UA})
and then takes the one-level approximation,
the widths are given as
\be
\Gamma_{\lambda c}(E) = 2 P_c(E) \gamma_{\lambda c}^2.
\ee
Thus, the formal parameters for the 
reduced widths of 
the present
formalism are
the same as the observed ones,
\be
\gamma_{\lambda c}^{\text{obs}} = \gamma_{\lambda c}.
\label{gg}\ee
This simple relation should be compared with
\be
\gamma_{\lambda c}^{\text{obs}}
=\check{\gamma}_{\lambda c}/ \sqrt{
1+ \sum_{c'} \check{\gamma}_{\lambda c'}^2 S_{\lambda c'}'},
\label{gamma_conv}\ee
where 
$\check{\gamma}_{\lambda c}$ are the formal reduced width parameters
in the conventional R-matrix theory.

I note that
the alternate parametrization obtained by Brune~\cite{brune}
has the same off-diagonal terms of $A^{-1}$ with
Eq.(\ref{Ainv2}).
But the diagonal elements are different,
and the values corresponding to $\J_{\lambda\lambda}$
used in Ref.~\cite{brune} are 1 
instead of those from Eq.(\ref{Kll}).
Consequently,
Brune's reduced width parameters 
are the same as those of the conventional R-matrix theory,
$\check{\gamma}_{\lambda c}$,
which are subject to the nonlinear relation 
described in Eq.(\ref{gamma_conv}).


\section{${}^{12}\text{C}+p$ elastic scattering}

To demonstrate 
the performance of the new R-matrix theory developed here,
I developed a simple Mathematica code~\cite{rx-code}
which calculates
the collision matrices and the differential cross-sections
of nuclear reactions
based on Eqs.(\ref{Blc},\ref{Kll},\ref{Ainv2}).
Since there is no need for conversions between the formal and the observed parameters,
a substantial simplification could be achieved.
The code is then applied to 
describe
${}^{12}\text{C}+p$ elastic scattering,
for which accurate experimental data are available~\cite{exp-data}.
For a detailed discussion of the process 
in connection with 
the (conventional) R-matrix theory, see Ref.~\cite{descouvemont}.

At low energy, this process is dominated by the three low-lying resonances
of ${}^{13}\text{N}$:
$J^\pi= 1/2^+$ at 0.421 MeV, $3/2^-$ at 1.559 MeV, and $5/2^+$ at 1.604 MeV,
where the resonance energies are 
the center-of-mass energies 
of the
${}^{12}\text{C}+p$ system.
I use the values of the {\em observed} parameters given in Ref.~\cite{descouvemont}
for my formal parameters,
which are listed in Table~1.
There is little dependence on the channel radius, 
which was chosen as $a_c=5$ fm for calculations.

\begin{table}[bp]
\centering
\caption{R-matrix parameters for ${}^{12}\text{C}+p$ elastic scattering with $a_c=5$ fm.
$E_\lambda$ are 
the center-of-mass energies of the
${}^{12}\text{C}+p$ system.
$E_\lambda$ and $\Gamma_{\lambda c}$ are from
Ref.~\cite{descouvemont},
and $\gamma_{\lambda c}$ are the corresponding reduced width parameters.
}
\label{t1}
\begin{tabular}{|c|ccc|}
\noalign{\smallskip}\noalign{\smallskip}
\hline
$J^\pi$ & $E_\lambda\,[\text{MeV}]$ & $\Gamma_{\lambda c}\,[\text{keV}]$ & $\gamma_{\lambda c}^2\,[\text{MeV}]$ \\
\hline
$1/2^+$ & 0.427 & 32.5 & 0.569 \\
\hline
$3/2^-$ & 1.559 & 51.4 & 0.0835 \\
\hline
$5/2^+$  & 1.604 & 48.1 & 0.414 \\
\hline
\end{tabular}
\end{table}

For two particular center-of-mass angles,
$\theta=89.1^\circ$ and $146.9^\circ$,
the resulting differential cross-sections 
with respect to the proton energy in the lab frame
are drawn 
in Fig.~1.
The figure shows that
the code with the newly developed R-matrix theory
reproduces nicely the experimental data~\cite{exp-data}.
\begin{figure}
\centering
  \includegraphics[width=0.4\linewidth]{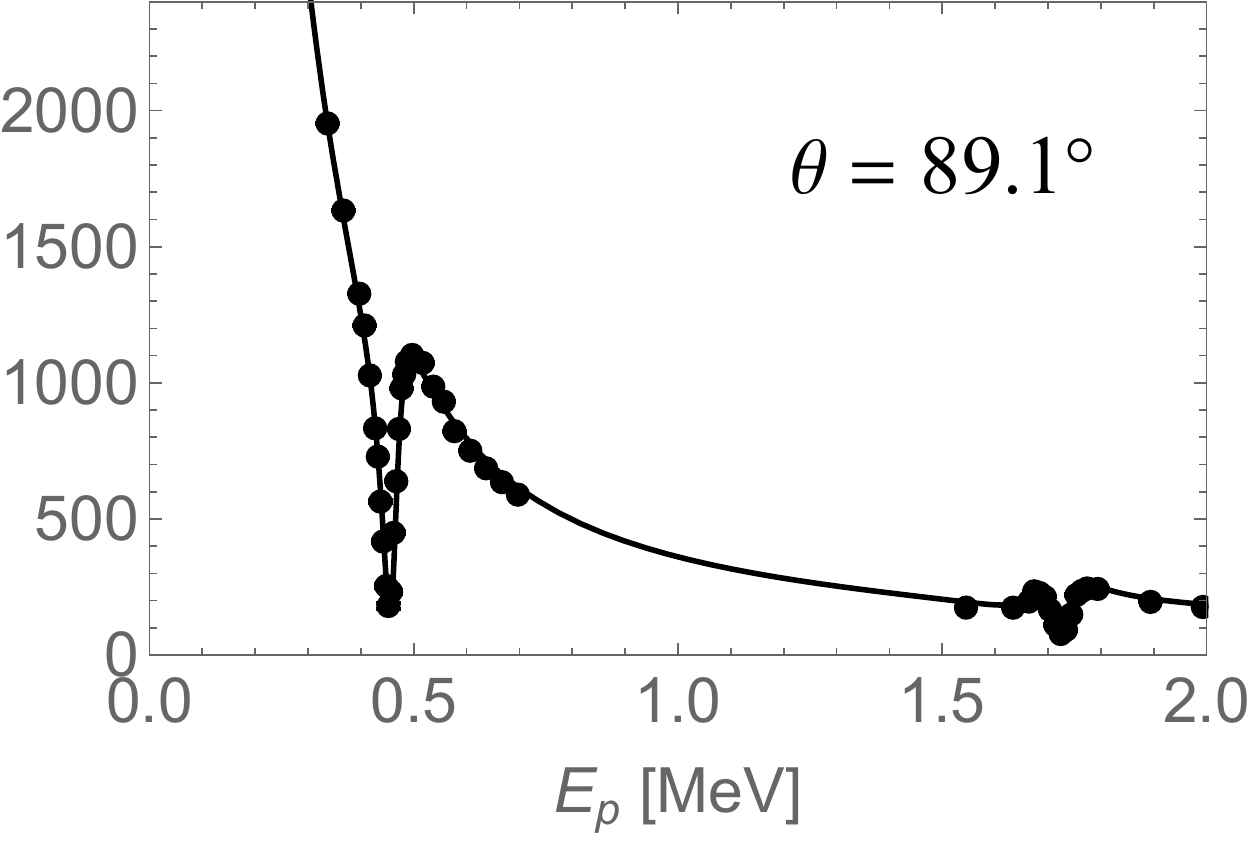}
%
  \includegraphics[width=0.4\linewidth]{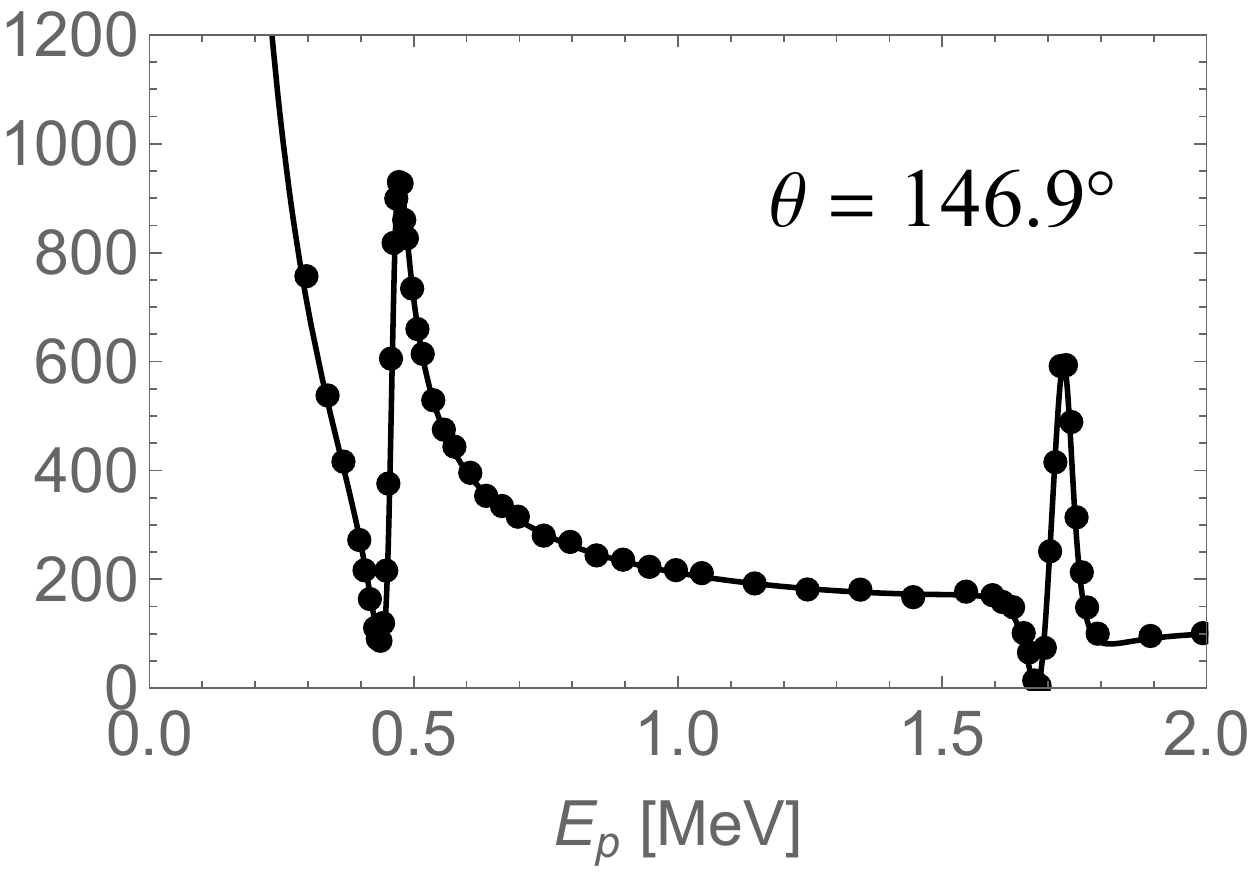}
\caption{
${}^{12}\text{C}+p$ elastic scattering cross-sections (in mb/sr)
with respect to the incident proton energy in the lab frame
for the center-of-mass angle $\theta=89.1^\circ$ (left panel)
and $146.9^\circ$ (right panel).
The experimental data are from Ref.~\cite{exp-data} (closed circles).
}
\end{figure}

In Table~2, the parameters of the conventional 
R-matrix theory are compared with those of 
the present
theory
for 
the first $J^\pi=1/2^+$ resonance in
${}^{12}\text{C}+p$,
varying the channel radius from 4 fm to 7 fm.
The observed position and the width of the resonance 
are set to 
$E_R=0.42$ MeV and $\Gamma_R=32$ keV~\cite{descouvemont}.
While the $E_1$ of the present theory is the same as the 
input value for the observed
pole position $E_R$ by construction, the formal parameter
$E_{1,\text{formal}}$ in the conventional theory
has a strong dependence on the channel radius.
The table also shows that $\gamma_1^2$ of the present theory
agrees well with $\gamma_{1,\text{observed}}^2$ in Ref.~\cite{descouvemont},
which is the intended outcome of this work.
The table also shows that 
these observed parameters are less
dependent on $a_c$
than the formal parameters.

\begin{table}
\centering
\caption{R-matrix parameters for the
first $1/2^+$ resonance 
($E_R=0.42\ \text{MeV}$ and $\Gamma_R=32\ \text{keV}$)
in ${}^{12}\text{C}+p$ 
elastic scattering (in MeV). 
The parameters of the conventional R-matrix theory
are from Table~10 of Ref.~\cite{descouvemont}.
}
\label{t2}
\begin{tabular}{|c|cccc|}
\noalign{\smallskip}\noalign{\smallskip}
\hline
$a_c\, [\text{fm}]$ & $4$ & $5$ & $6$ & $7$ \\
\hline
$\gamma_{1,\text{observed}}^2$ (Ref.~\cite{descouvemont}) & 1.089 & 0.592 & 0.353 & 0.227 \\
$\gamma_{1,\text{formal}}^2$ (Ref.~\cite{descouvemont})& 3.083 & 1.157 & 0.569 & 0.323 \\
$\gamma_{1}^2$ (this work) & 1.087 & 0.591& 0.353 & 0.226 \\
\hline
$E_{1,\text{formal}}$(Ref.~\cite{descouvemont}) & $-2.152$ & $-0.614$ & $-0.110$ & 0.113 \\
$E_1$ (this work) & $0.42$ & $0.42$ & $0.42$ & 0.42 \\
\hline
\end{tabular}
\end{table}

\section{Discussions}

The conventional R-matrix theory 
has the problem of having {\em formal} parameters that are different from the observed ones,
and thus requiring non-trivial conversions between the two sets of parameters.
As discussed in the text,
this drawback is
a consequence of requiring
orthogonality
of the basis states for the Hilbert space in the internal region of the R-matrix theory.
%
However, the orthogonality is not a necessary condition 
for the basis.
By exploiting the additional freedom that can be achieved when 
the orthogonality condition is released,
I have developed a new R-matrix theory
that has no distinction between the two sets of parameters.

In this theory, 
the boundary condition parameters are
allowed to be 
level-dependent
and
adjusted to make 
the {\em formal} parameters $E_\lambda$
identical to the {\em observed} pole positions.
That is, I assigned $B_{\lambda c}$ to
the shift factor of channel $c$ at $E_\lambda$,
see Eq.(\ref{Blc}), which makes
the secular equation Eq.(\ref{secular}) vanish at that energy.
Recalling that the observed pole-positions $E_\lambda^{\text{obs}}$ are 
defined to be
the zeroes of the secular equation,
one sees that the imposed boundary condition
leads to $E_\lambda = E_\lambda^{\text{obs}}$.




In addition,
there is another freedom in the normalization of the diagonal elements $J_{\lambda\lambda}$,
which correspond to the square of the norm of the basis states.
%
%
By selecting the normalization factor of the basis states
as given in Eq.(\ref{Kll}),
which can be viewed as a natural extrapolation of the off-diagonal elements,
I could derive the {\em formal} reduced width parameters 
to be the same as the {\em observed} ones as well,
$\gamma_{\lambda c}^{\text{obs}} = \gamma_{\lambda c}$.
As a result, there are no {\em formal} parameters 
which are
different from the 
{\em observed} ones in the present formalism.

As a demonstration, 
I tested a computation code based on 
the developed R-matrix theory, 
where the trial case was
the elastic scattering of protons on ${}^{12}\text{C}$.
The code required only the resonance data as input
and did not invoke any transformations of parameters.
The code was able to reproduce
the experimental differential cross sections quite well.


%
\section*{Acknowledgments}
I thank E.J. In, S.-W. Hong and J. Park for valuable discussions.
This work was supported by the Institute for Basic Science (IBS-R031-D1).

\end{document}